\documentclass{optica-article}
\journal{josab}
\articletype{Research Article}

\begin{document}
\title{Enhancing the quantum entanglement and EPR steering of a coupled optomechanical system with a squeezed vacuum field}
\author{Shao-Xiong Wu,\authormark{1,2,*}, Cheng-Hua Bai,\authormark{2} Gang Li,\authormark{1,3} Chang-shui Yu,\authormark{4,\dag}, and Tiancai Zhang\authormark{1,3,\ddag}}
\address{\authormark{1} State Key Laboratory of Quantum Optics and Quantum Optics Devices, and Institute of Opto-Electronics, Shanxi University, Taiyuan 030006, China\\
\authormark{2} School of Semiconductor and Physics, North University of China, Taiyuan 030051, China\\
\authormark{3} Collaborative Innovation Center of Extreme Optics, Shanxi University, Taiyuan 030006, China\\
\authormark{4} School of Physics, Dalian University of Technology, Dalian 116024, China}
\email{\authormark{*}sxwu@nuc.edu.cn}
\email{\authormark{\dag}ycs@dlut.edu.cn}
\email{\authormark{\ddag}tczhang@sxu.edu.cn}

\begin{abstract*}
Quantum entanglement and Einstein-Podolsky-Rosen (EPR) steering are valuable resources in quantum information processing. How to enhance the quantum entanglement and EPR steering of coupled optomechanical systems with a weak squeezed vacuum field are studied when the displacement of detuning induced by the mechanical mode is considered. Compared with the condition that the system interacts with a vacuum environment, the quantum entanglement and EPR steering are stronger when the squeezed vacuum field is applied. A squeezed vacuum field with a large degree is not beneficial to enhance the quantum entanglement and EPR steering. Rather than the squeezing parameter of the squeezed vacuum field, the reference phase plays a vital role in this model.
\end{abstract*}

\section{Introduction}
In the famous Gedanken experiment, Einstein, Podolsky, and Rosen (EPR) presented the quantum correlation problem of two-body systems \cite{ea35}. As a discussion of the EPR paradox, Schr\"{o}dinger envisages the concept of quantum entanglement, which is the critical resource in quantum information processing and possesses a solid foundation both theoretically and experimentally \cite{se35,hr09,okcf18}. To control the quantum entanglement or allow one system to steer or pilot another system, EPR steering is proposed at the same time \cite{se35}, which is stronger than the quantum entanglement and weaker than the Bell nonlocality under the hierarchy of quantum nonlocality \cite{rmd89,rmd09,wsp11}.

Realizing strong quantum nonlocality (including quantum entanglement and EPR steering) in continuous variable or discrete version and exploring the novel nonclassical effects of quantum systems are significant issues in quantum optics and quantum information. With the development of experimental technology, especially the micro/nano manufacturing technology, optomechanical systems have become ideal and sensitive platforms for testing and exploring the fundamental theory of quantum mechanics \cite{am14,mi18}, and the quantum effects and nonlocality of optomechanical systems have been investigated extensively. Stationary entanglement between the cavity mode and mechanical mode can be generated using radiation pressure \cite{vd07,gc08}; various methods of generating entanglement have been proposed, for example, by modulation in optical machinery \cite{ma09}, reservoir engineering \cite{wyd13}, dissipation process \cite{hx15}, cross-Kerr nonlinearity \cite{cs17}, pump optical degenerate parametric amplifier \cite{hcs20}, coupling pumped auxiliary cavity \cite{ldg21}, using optomechanical array \cite{hxz21}, pump modulation double-cavity \cite{zwj21}, amplitude-modulated pump field \cite{bch21}, and quantum interference \cite{wf22}. Besides the quantum entanglement of the optical and mechanical modes, the squeezing and cooling of the mechanical modes have also received attentions, such as, utilizing modulation of the external driving field \cite{ljq11}, dynamic dissipative cooling \cite{lyc13}, electromagnetically-induced-transparency-like cooling \cite{gy14}, Duffing nonlinearity \cite{lxy15pra,hx19}, Lyapunov control \cite{xb20}, and pump modulation \cite{bch20}.

In addition to quantum entanglement, EPR steering is another common measure of quantum nonlocality, which shows intrinsic asymmetry. The properties and applications of EPR steering in optomechanical systems have become essential topics in recent years, for example, the operational connection between Gaussian steerability and secure teleportation \cite{hq15,sfx17}, phase control of quantum entanglement and EPR steering in three-mode optomechanical systems \cite{zj15}, EPR steering mediated by mechanical oscillators \cite{th15}, via atomic coherence \cite{ps07,zw17}, driven by two-tone lasers \cite{gq23} or two four-tone lasers \cite{lcg20}, utilizing interference effects induced by closed-loop coupling \cite{zs19}, EPR steering in hybrid cavity magnomechanical systems \cite{lj17,yzb23,zw22}, and so on.

In quantum physics processes, the quantum squeezed states are valuable quantum resources, such as enhancing the single-photon optomechanical coupling strength by using nonlinear interaction between squeezed cavity mode and mechanical mode \cite{lxy15}, achieving optical nonreciprocity on-chip via quantum squeezing driving laser \cite{tl22}, inducing superradiant phase transition by optical parametric amplification squeezed process \cite{zcj20}, and advancing quantum-squeezing-enhanced weak-force sensing via nonlinear optomechanical resonator \cite{zw20}. Motivated by these works, we investigate the quantum entanglement and EPR steering of coupled optomechanical systems with a weak squeezed vacuum field in this paper. What are the squeezed vacuum field's constructive or destructive effects on quantum entanglement and EPR steering? How can the adverse impact be eliminated and the quantum nonlocality be enhanced compared with the vacuum environment? These are essential issues and deserve to be studied systematically.

The structure of this paper is as follows. In Sec. \ref{sec2}, a optomechanical model that contains two optical modes and one mechanical mode is introduced, and the linearized effective Hamiltonian is obtained. In Sec. \ref{sec3}, we study the dynamics of quadrature fluctuation operators and the system's stability. In Sec. \ref{sec4}, we analyze the effects of a squeezed vacuum field on quantum entanglement and EPR steering, and how to enhance quantum nonlocality is expressed. The discussion and conclusion are given at the end.

\section{The model and the linearized effective Hamiltonian}\label{sec2}
\begin{figure}[t]
\centering
\includegraphics[width=0.85\columnwidth]{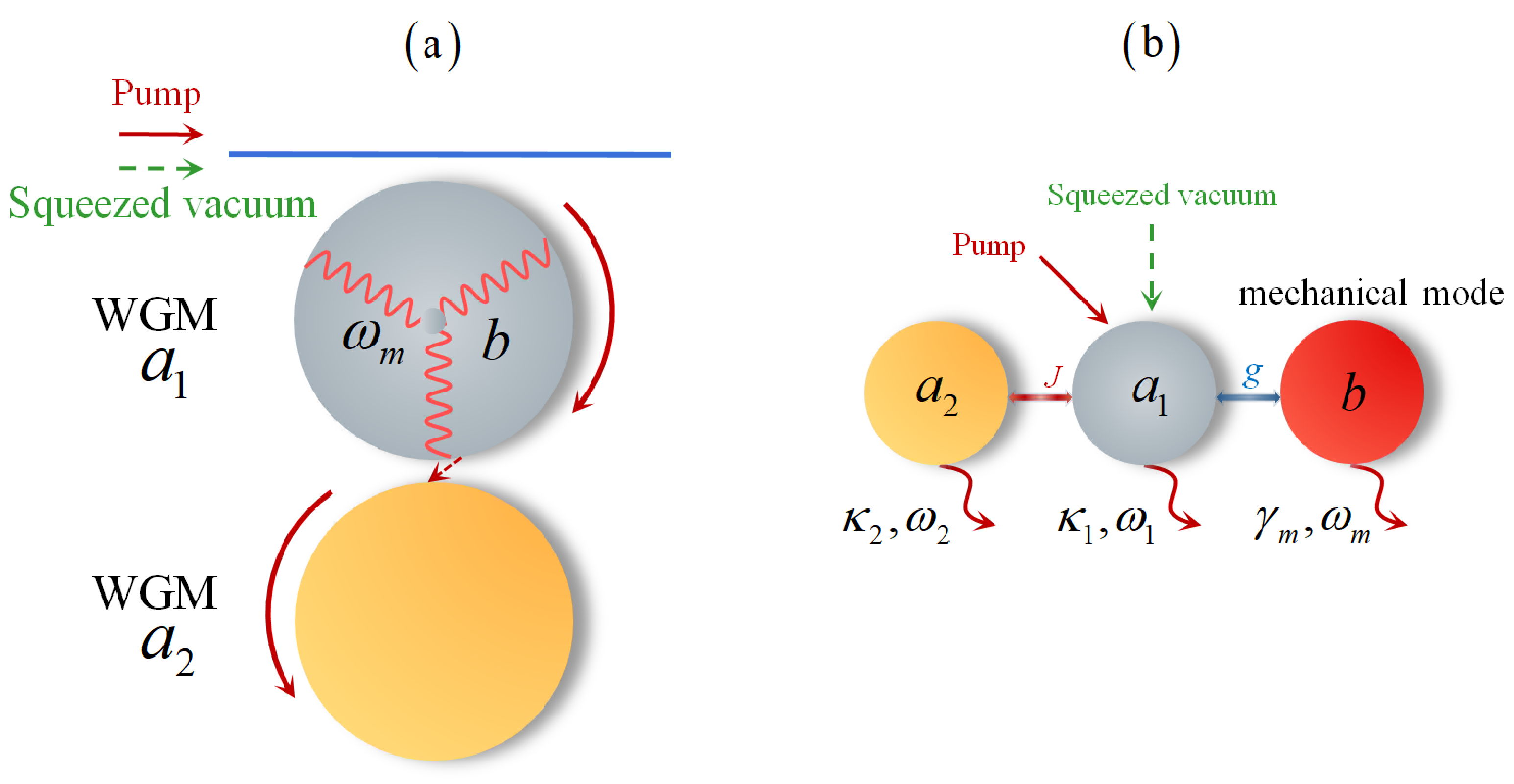}
\caption{Schematic diagrams of the model. (a) The model consists of two coupled whispering-gallery-mode (WGM) resonators and one optomechanical resonator, and is driven by a pump laser and a weak squeezed vacuum field. (b) The WGM modes $a_1$ and $a_2$ are directly coupled with strength $J$, their frequencies are $\omega_1$, $\omega_2$, and the decay rates are $\kappa_1$ and $\kappa_2$. The WGM mode $a_1$ and mechanical mode $b$ (frequency $\omega_m$ and decay rate $\gamma_m$) are coupled with the single-photon coupling strength $g$.}
\label{fig1}
\end{figure}

As depicted in Fig. \ref{fig1}, the considered optomechanical model comprises two coupled whispering-gallery-mode (WGM) resonators and one mechanical resonator and is driven by a pump laser and a fragile squeezed vacuum field. The optomechanical WGM resonator consists of the optical mode $a_1$ with frequency $\omega_1$ and decay rate $\kappa_1$, and the mechanical mode $b$ with frequency $\omega_m$ and decay rate $\gamma_m$, where the single-photon optomechanical coupling strength between the modes $a_1$ and $b$ is $g$. The frequency of the WGM resonator $a_2$ is $\omega_2$, and the corresponding decay rate is $\kappa_2$. The coupling strength between the WGM modes $a_1$ and $a_2$ is $J$. The Hamiltonian (in units of $\hbar$) of the whole optomechanical system can be given as
\begin{align}
H=&\omega_1a_1^{\dagger}a_1+\omega_2a_2^{\dagger}a_2+\omega_mb^{\dagger}b-g a_1^{\dagger}a_1(b+b^{\dagger})\notag\\
&+J(a_1^{\dagger}a_2+a_1a_2^{\dagger})+iE(a_1^{\dagger}e^{-i\omega t}-a_1e^{i\omega t}),
\label{eq:Hyuanshi}
\end{align}
where $E$ is the amplitude of external coherent pump laser with frequency $\omega$. The intensity of the squeezed vacuum field is assumed to be very faint, so it only affects the fluctuation of the cavity mode $a_1$, and can be treated as a noise term when dealing with the evolution of the quantum system. In the frame rotating under the pump laser frequency $\omega$, i.e., $U^{\dagger}HU-iU^{\dagger}\dot{U}$ with $U=\exp[-i\omega t(a_1^{\dagger}a_1+a_2^{\dagger}a_2)]$, the Hamiltonian can be rewritten as
\begin{align}
H_{r}=&\Delta_1a_1^{\dagger}a_1+\Delta_2a_2^{\dagger}a_2+\omega_{m}b^{\dagger}b-ga_1^{\dagger}a_1(b+b^{\dagger})\notag\\
&+J(a_1^{\dagger}a_2+a_1a_2^{\dagger})+iE(a_1^{\dagger}-a_1),
\end{align}
where $\Delta_1=\omega_1-\omega$ is the detuning between the WGM mode $a_1$ and the pump laser, and $\Delta_2=\omega_2-\omega$.

Because the physical systems inevitably interact with their surroundings, the environment's noise should be considered in the evolution. Taking into account the dissipation and noise of WGM modes $a_1$, $a_2$ and mechanical mode $b$, the dynamics of the system are governed by the quantum Langevin equations as follows \cite{gcw00,smo97}:
\begin{align}
\dot{a_1}=&-i\Delta_1a_1-iJa_2+iga_1(b+b^{\dagger})+E -\frac{\kappa_1}{2}a_1+\sqrt{\kappa_1}a_1^{\text{in}},\notag\\
\dot{a_2}=&-i\Delta_2a_2-iJa_1-\frac{\kappa_2}{2}a_2+\sqrt{\kappa_2}a_2^{\text{in}},\notag\\
\dot{b}=&-i\omega_mb+iga_1^{\dagger}a_1-\frac{\gamma_m}{2}b+\sqrt{\gamma_m}b^{\text{in}},
\label{eq:qle1}
\end{align}
where $a_{1}^{\text{in}}$, $a_{2}^{\text{in}}$ and $b^{\text{in}}$ are the zero-mean input WGM mode $a_1$ squeezed vacuum field noise operator, WGM mode $a_2$ vacuum noise operator and mechanical mode $b$ thermal noise, respectively. Under the Markovian approximation, the nonzero correlation functions of the squeezed vacuum field operator $a_1^{\text{in}}$ are
\begin{align}
\langle a_1^{\text{in}\dagger}(t)a_1^{\text{in}}(t')\rangle&=N\delta(t-t'),\notag\\
\langle a_1^{\text{in}}(t)a_1^{\text{in}\dagger}(t')\rangle&=(N+1)\delta(t-t'),\notag\\
\langle a_1^{\text{in}}(t)a_1^{\text{in}}(t')\rangle&=M\delta(t-t'),\notag\\
\langle a_1^{\text{in}\dagger}(t)a_1^{\text{in}\dagger}(t')\rangle&=M^{*}\delta(t-t'),
\label{eq:zaosheng1}
\end{align}
where $N=\sinh^{2}r$ and $M=e^{i\theta}\sinh r\cosh r$. The correlation functions of noise operators $a_2^{\text{in}}$ and $b^{\text{in}}$ meet:
\begin{align}
\langle a_2^{\text{in}\dagger}(t)a_2^{\text{in}}(t')\rangle&=n\delta(t-t'),\notag\\
\langle a_2^{\text{in}}(t)a_{2}^{\text{in}\dagger}(t')\rangle&=(n+1)\delta(t-t'),\notag\\
\langle b^{\text{in}\dagger}(t)b^{\text{in}}(t')\rangle&=m\delta(t-t'),\notag\\
\langle b^{\text{in}}(t)b^{\text{in}\dagger}(t')\rangle&=(m+1)\delta(t-t'),
\label{eq:zaosheng2}
\end{align}
where $n$ and $m$ are the thermal mean photon/phonon numbers of noise for WGM mode $a_2$ and mechanical mode $b$, respectively. It is reasonable that $a_2^{\text{in}}$ is treated as vacuum noise due to the tremendous difference between the optical and mechanical frequencies.

The quantum Langevin equations in Eq. (\ref{eq:qle1}) can be addressed with the standard linearization technology under strong external coherent pump, and the operators of WGM modes $a_1$, $a_2$ and mechanical mode $b$ can be rewritten as the sum of the steady-state mean values and the quantum fluctuation operators, i.e., $o=\langle o\rangle+\delta o$ $(o=a_1,a_2,b)$. The differential equations for the steady-state mean values are
\begin{align}
\langle\dot{a_1}\rangle &=-i\Delta_1'\langle a_1\rangle-iJ\langle a_2\rangle+E-\frac{\kappa_1}{2}\langle a_1\rangle,\notag\\
\langle\dot{a_2}\rangle &=-i\Delta_2\langle a_2\rangle-iJ\langle a_1\rangle-\frac{\kappa_2}{2}\langle a_2\rangle,\notag\\
\langle\dot{b}\rangle &=-i\omega_{m}\langle b\rangle+ig|\langle a_1\rangle|^2 -\frac{\gamma_m}{2}\langle b\rangle.
\end{align}
The linearized quantum Langevin equations for the quantum fluctuation operators can be given in the following form:
\begin{align}
\dot{\delta a_1}=&-i\Delta'\delta a_1-iJ\delta a_2+iG(\delta b+\delta b^{\dagger})-\frac{\kappa_1}{2}\delta a_1+\sqrt{\kappa_1}a_1^{\text{in}},\notag\\
\dot{\delta b}=&-i\omega_{m}\delta b+i(G^*\delta a_1+G\delta a_1^{\dagger})-\frac{\gamma_m}{2}\delta b+\sqrt{\gamma_m}b^{\text{in}},\notag\\
\dot{\delta a_2}=&-i\Delta_2\delta a_2-iJ\delta a_1-\frac{\kappa_2}{2}\delta a_2+\sqrt{\kappa_2}a_2^{\text{in}}.
\label{eq:qle2}
\end{align}
Therefore, the linearized effective Hamiltonian is arrived at:
\begin{align}
H_{\text{eff}}=&\Delta_1'\delta a_1^{\dagger}\delta a_1+\Delta_2\delta a_2^{\dagger}\delta a_2+J(\delta a_1\delta a_2^{\dagger}+\delta a_1^{\dagger}\delta a_2)\notag\\
&+\omega_m\delta b^{\dagger}\delta b-(G^*\delta a_1+G\delta a_1^{\dagger})(\delta b^{\dagger}+\delta b).
\label{eq:linH}
\end{align}
The effective detuning $\Delta_1'$ and the effective coupling strength $G$ are determined by
\begin{align}
\Delta_1'=\Delta_1-2g\text{Re}\langle b\rangle,G=g\langle a_1\rangle,
\label{eq:Gcanshu}
\end{align}
with the coefficients
\begin{gather}
\text{Re}\langle b\rangle=\frac{g|\langle a_1\rangle|^2\omega_m}{\omega_m^2+\left(\frac{\gamma_m}{2}\right)^2},\notag\\
\langle a_1\rangle=\frac{E\left(i\Delta_2+\frac{\kappa_2}{2}\right)}
{J^2+\left(i\Delta_1'+\frac{\kappa_1}{2}\right)\left(i\Delta_2+\frac{\kappa_2}{2}\right)}.
\end{gather}
It is easy to find that $\Delta_1'$ and $G$ are independent, and $G$ is generally a complex number. Therefore, the red-detuned driving assumption $\Delta_1'=\omega_m$ and the rotating-wave approximation are not employed. In the following, we will analyze the effects of the squeezed vacuum field on the quantum nonlocality (including quantum entanglement and EPR steering) in the resolved sideband regime ($\kappa_1<\omega_m$) when the displacement of detuning $2g\text{Re}\langle b\rangle$ induced by the mechanical mode is considered.

\section{Dynamics of the quadrature fluctuation operators and the stability of system}\label{sec3}
To investigate the dynamics of the optomechanical system, the position and momentum quadrature fluctuation operators and quadrature noise operators are introduced as
\begin{align}
X_{o}=\frac{o+o^{\dagger}}{\sqrt{2}},~& Y_{o}=\frac{o-o^{\dagger}}{i\sqrt{2}},\notag\\
X_{o^{\text{in}}}=\frac{o^{\text{in}}+o^{\text{in}\dagger}}{\sqrt{2}},~& Y_{o^{\text{in}}}=\frac{o^{\text{in}}-o^{\text{in}\dagger}}{i\sqrt{2}},
\label{eq:zhengjiaosuanfu}
\end{align}
with $o=\delta a_1$, $\delta a_2$, $\delta b$, and $o^{\text{in}}=a_1^{\text{in}}$, $a_2^{\text{in}}$, $b^{\text{in}}$. The vectors of the quadrature fluctuation and noise operators in Eq. (\ref{eq:zhengjiaosuanfu}) are set as $R(t)=[X_{\delta a_1}, Y_{\delta a_1}, X_{\delta a_2}, Y_{\delta a_2}, X_{\delta b}, Y_{\delta b}]^{T}$ and $N(t)=[X_{a_1^{\text{in}}}, Y_{a_1^{\text{in}}}, X_{a_2^{\text{in}}}, Y_{a_2^{\text{in}}}, X_{b^{\text{in}}}, Y_{b^{\text{in}}}]^{T}$, respectively. Therefore, the linearized quantum Langevin equations for the quantum fluctuation operators in Eq. (\ref{eq:qle2}) can be reexpressed as
\begin{align}
\frac{dR(t)}{dt}=\mathcal{M}(t)R(t)+N(t),
\end{align}
where the drift matrix $\mathcal{M}(t)$ is
\begin{align}
\mathcal{M}(t)=\left[\begin{array}{cccccc}
-\frac{\kappa_1}{2} & \Delta_1' & 0 & J & -2G_y & 0\\
-\Delta_{1}' & -\frac{\kappa_1}{2} & -J & 0 & 2G_x & 0\\
0 & J & -\frac{\kappa_2}{2} & \Delta_2 & 0 & 0\\
-J & 0 & -\Delta_2 & -\frac{\kappa_2}{2} & 0 & 0\\
0 & 0 & 0 & 0 & -\frac{\gamma_m}{2} & \omega_m\\
2G_x & 2G_y & 0 & 0 & -\omega_m & -\frac{\gamma_m}{2}
\end{array}\right]
\label{eq:shifitmatrix}
\end{align}
with coefficients $G_x=\text{Re}G$ and $G_y=\text{Im}G$ being the real and imaginary parts of the effective coupling strength $G$.

The dynamic evolution of the quadrature fluctuation operators is governed by the linearized effective Hamiltonian $H_{\text{eff}}$ in Eq. (\ref{eq:linH}), and due to the Gaussian nature of the quantum noise, the system of the quadrature fluctuation operators forms a three-mode Gaussian state, which can be characterized by the $6\times 6$ covariance matrix $\mathcal{V}(t)$ with elements $\mathcal{V}_{kl}=\langle R_{k}R_{l}+R_{l}R_{k}\rangle/2$. For a steady state, its dynamics are determined by the Lyapunov equation:
\begin{align}
\mathcal{M}\mathcal{V}+\mathcal{V}\mathcal{M}^{T} +\mathcal{D}=0,
\label{eq:xiefangcha}
\end{align}
where $\mathcal{D}$ is the noise diffusion matrix with components $\mathcal{D}_{kl}=\langle N_{k}N_{l}+N_{l}N_{k}\rangle/2$. According to Eqs. (\ref{eq:zaosheng1}) and (\ref{eq:zaosheng2}), the noise diffusion matrix can be expressed in the following form
\begin{align}
\mathcal{D}=\mathcal{D}_{1}\oplus \mathcal{D}_{2}
\label{eq:zaoshengjuzhen}
\end{align}
with block matrices
\begin{align}
\mathcal{D}_{1} & =\frac{\kappa_{1}}{2}
\left[\begin{array}{cc}
2N+1+M+M^{*} & i(M^{*}-M)\\
i(M^{*}-M) & 2N+1-M-M^{*}
\end{array}\right],\notag\\
\mathcal{D}_{2} & =\text{diag}\left[\frac{\kappa_2}{2},\frac{\kappa_2}{2},\frac{\gamma_m}{2}(2m+1),\frac{\gamma_m}{2}(2m+1)\right].
\end{align}

For arbitrary two-mode Gaussian state among the three modes, its reduced $4\times 4$ covariance matrix can be rewritten in block matrix form as
\begin{align}
\mathcal{V}_{12}=\left[
                   \begin{array}{cc}
                     \mathcal{V}_1 & \mathcal{V}_c \\
                     \mathcal{V}_c^T & \mathcal{V}_2 \\
                   \end{array}
                 \right],
\end{align}
where $\mathcal{V}_1$, $\mathcal{V}_2$ and $\mathcal{V}_c$ are $2\times2$ subblock matrices corresponding to the interested modes 1, 2 and their correlated part, respectively. For Gaussian entanglement, a widely used metric is logarithmic Negativity \cite{vg02,ag04,pmb05}:
\begin{align}
E_N=\max\left\{0,-\ln2\eta^-\right\},
\end{align}
where $\eta^-=\sqrt{\Sigma-\sqrt{\Sigma^2-4\det\mathcal{V}_{12}}}/\sqrt{2}$ is the smallest symplectic eigenvalue of the partially transposed covariance matrix $\mathcal{V}_{12}$ with coefficient $\Sigma=\det \mathcal{V}_1+\det \mathcal{V}_2-2\det \mathcal{V}_{c}$. EPR steering, another broadly used quantum nonlocality measurement, occupies the intermediate hierarchy between quantum entanglement and Bell nonlocality, and its asymmetric form from mode 1 to 2 is defined \cite{ki15} as
\begin{align}
\mathcal{G}_{1\rightarrow2}= \max\left\{0,\frac{1}{2}\ln\frac{\det\mathcal{V}_1}{4\det\mathcal{V}_{12}}\right\},
\end{align}
and the EPR steering from mode 2 to 1 is defined as
\begin{align}
\mathcal{G}_{2\rightarrow1}= \max\left\{0,\frac{1}{2}\ln\frac{\det\mathcal{V}_2}{4\det\mathcal{V}_{12}}\right\}.
\end{align}

As a significant feature to distinguish from quantum entanglement, EPR steering exhibits obvious directivity. When $\mathcal{G}_{1\rightarrow2}>0$ and $\mathcal{G}_{2\rightarrow1}=0$, there is only Gaussian mode $1\rightarrow2$ one-way steering, and mode $1$ can steer mode $2$; however mode $2$ cannot steer mode $1$. And vice versa in the case that $\mathcal{G}_{2\rightarrow1}>0$ and $\mathcal{G}_{1\rightarrow2}=0$. From the quantum information theory, EPR steering can mean the ability that one individual system manipulate the state of another system by local measurement \cite{whm07,hq15}. Assuming the quantum entanglement distribution task between $A$ and $B$ who are far apart and do not trust each other. If $A$ can convince $B$ that they had shared a pair of entangled states under the condition that $B$ does not trust $A$, which can be considered that $A$ can steer the state of $B$, i.e., one-way steering $\mathcal{G}_{A\rightarrow B}$. EPR steering maybe more important resource than the quantum entanglement in the trustless quantum communication tasks.

\begin{figure}[t]
\centering
\includegraphics[width=1\columnwidth]{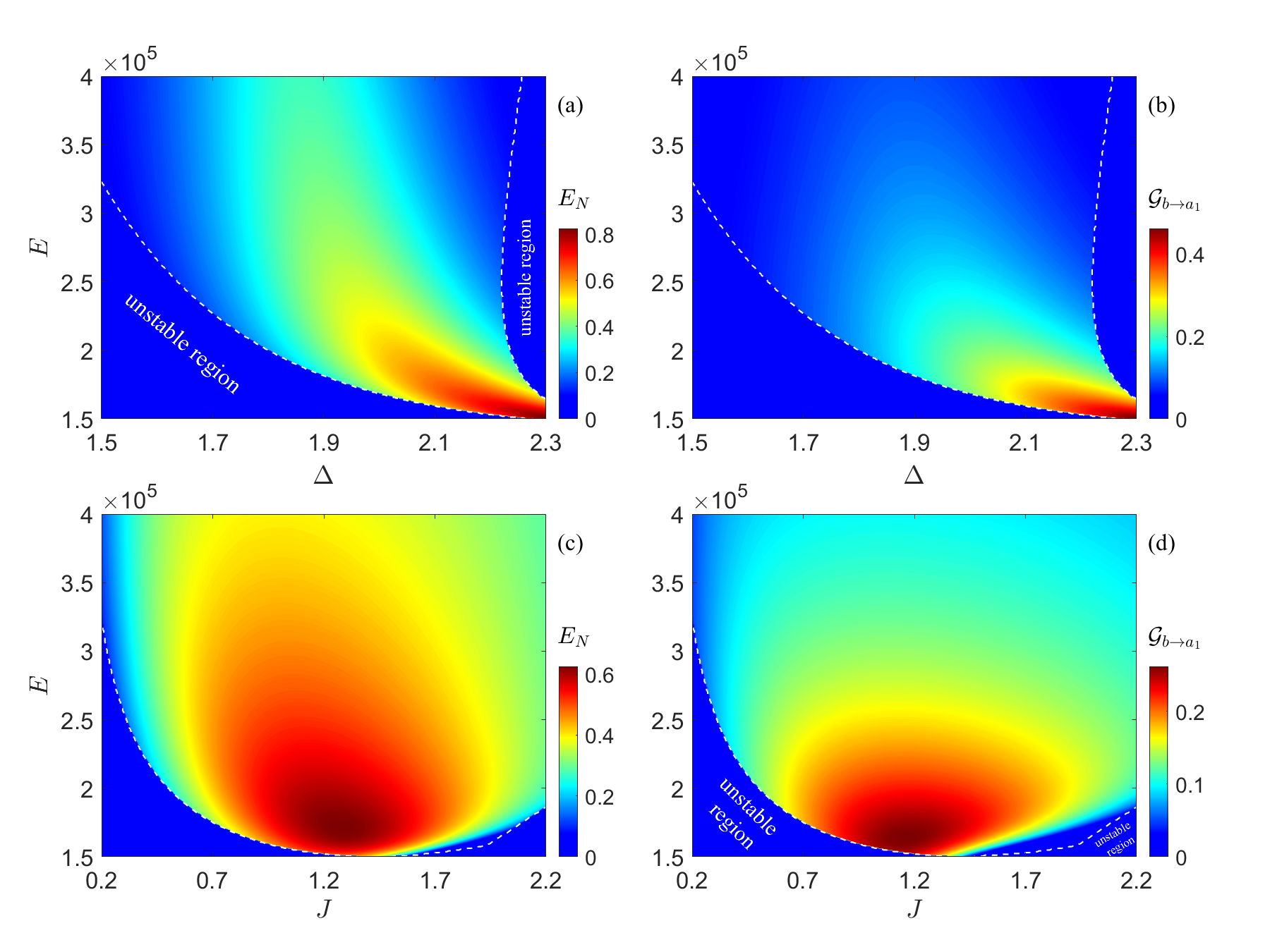}
\caption{The variations in quantum entanglement $E_N$ (a) and EPR steering $\mathcal{G}_{b\rightarrow a_1}$ (b) as functions of the detuning $\Delta$ and pump amplitude $E$ with the coupling strength $J=0.5$. $E_N$ (c) and $\mathcal{G}_{b\rightarrow a_1}$ (d) versus $J$ and $E$ with $\Delta=1.8$. In all panels, the decay rates of the WGM modes are set as $\kappa_1=\kappa_2=0.2$, the decay rate of the mechanical mode is $\gamma_m=10^{-5}$, and the single-photon coupling strength is $g=5.0\times10^{-5}$. All the parameters are in units of the mechanical resonator's frequency $\omega_m$.}
\label{fig2}
\end{figure}

According to the Routh-Hurwitz criterion \cite{dex87}, the nonlinear system is stable if and only if all eigenvalues of the shift matrix $\mathcal{M}(t)$ have negative real parts. In Fig. \ref{fig2}, we plot the quantum entanglement $E_N$ between the WGM mode $a_1$ and mechanical mode $b$ and the EPR steering $\mathcal{G}_{b\rightarrow a_1}$ from mode $b$ to $a_1$, and the stability of the system is shown at the same time where the deep blue regions represent the unstable regime and the cross bound between the stable and unstable regimes are marked by white dashed line. Without loss of generality, the qualities and frequency of the WGM resonators $a_1$ and $a_2$ are assumed as to be identical, the decay rates are set as $\kappa_1=\kappa_2=0.2$, and the detuning between the pump laser and the WGM modes $a_1$ and $a_2$ are $\Delta_1=\Delta_2=\Delta$. The decay rate of the mechanical oscillator is chosen as $\gamma_m=10^{-5}$, and the single-photon optomechanical coupling strength is $g=5.0\times 10^{-5}$. All the parameters are in units of the frequency of mechanical oscillator $\omega_m$.  Because the noise diffusion matrix $\mathcal{D}$ does not affect the system's stability, the squeezing parameter and mean thermal phonon number are chosen as $r=0$ and $m=0$. The EPR steering $\mathcal{G}_{b\rightarrow a_1}$ shows similar behavior with quantum entanglement $E_N$ and is weaker than $E_N$, which is governed by the hierarchy of nonlocality. Moreover, it can be checked that the EPR steering $\mathcal{G}_{a_1\rightarrow b}$ from mode $a_1$ to $b$ is zero in all stable regimes, which shows the intrinsic asymmetry feature between $\mathcal{G}_{b\rightarrow a_1}$ and $\mathcal{G}_{a_1\rightarrow b}$. Hence, mechanical mode $b$ can steer WGM mode $a_1$ with suitable system parameters. However, mode $a_1$ cannot steer mode $b$.

According to the effective Hamiltonian $H_{\text{eff}}$ (\ref{eq:linH}), the interaction Hamiltonian between modes $a_1$ and $b$ is $(G^*\delta a_1\delta b+G\delta a_1^\dagger\delta b^\dagger)+(G^*\delta a_1\delta b^\dagger+G\delta a_1^\dagger\delta b)$. The first part represents the generation process of a two-mode squeezed state between the WGM mode $a_1$ and mechanical mode $b$ so that this process will have quantum entanglement. The second beam-splitter part expresses the state transfer process between modes $a_1$ and $b$. The effective coupling strength $G$ and its complex conjugate $G^*$, and the maximal quantum entanglement and EPR steering are jointly determined by the system parameters, which are shown in Fig. \ref{fig2}. Based on the shift matrix $\mathcal{M}(t)$ in Eq. (\ref{eq:shifitmatrix}), the detuning parameters $\Delta_1'$ and $\Delta_2$ play equivalent roles under the dynamics, which are of the same order of magnitude as the mechanical mode's frequency $\omega_m$. Taking into account the displacement of detuning $2g\text{Re}\langle b\rangle$ induced by the mechanical mode into $\Delta'$, we give the contour plots of $E_N$ and $\mathcal{G}_{b\rightarrow a_1}$ as functions of the original detuning $\Delta$ between the WGM mode $a_1$ and pump laser and the pump amplitude $E$ in Fig. \ref{fig2} (a) and (b) with the coupling strength $J=0.5$.

There is another state transfer path between the WGM modes $a_1$ and $a_2$ through the beam-splitter interaction $J(\delta a_1\delta a_2^\dagger+\delta a_1^\dagger\delta a_2)$. Intuitively, the quantum nonlocality between the modes $a_1$ and $b$ is stronger when the coupling strength $J$ vanishes. However, the parameter $J$ can significantly change the system's stability. In Fig. \ref{fig2}(c) and (d), the variations in quantum entanglement $E_N$ and EPR steering $\mathcal{G}_{b\rightarrow a_1}$ with the coupling strength $J$ and pump amplitude $E$ are depicted under the detuning $\Delta=1.8$. From Fig. \ref{fig2}, one can find that both $E_N$ and $\mathcal{G}_{a_1\rightarrow b}$ first increase and then decrease with  increasing pump amplitude. Although the areas of the stable region become larger, both the quantum entanglement $E_N$ and EPR steering $\mathcal{G}_{b\rightarrow a_1}$ are weaker. Therefore, it is advantageous to choose a moderate pump intensity to generate strong, stable quantum nonlocality. In the next section, we will investigate the effects of a squeezed vacuum field on the quantum entanglement and EPR steering and analyze how to enhance the quantum nonlocality utilizing this weak squeezed vacuum field in the stable regime.

\begin{figure}[t]
\centering
\includegraphics[width=0.5\columnwidth]{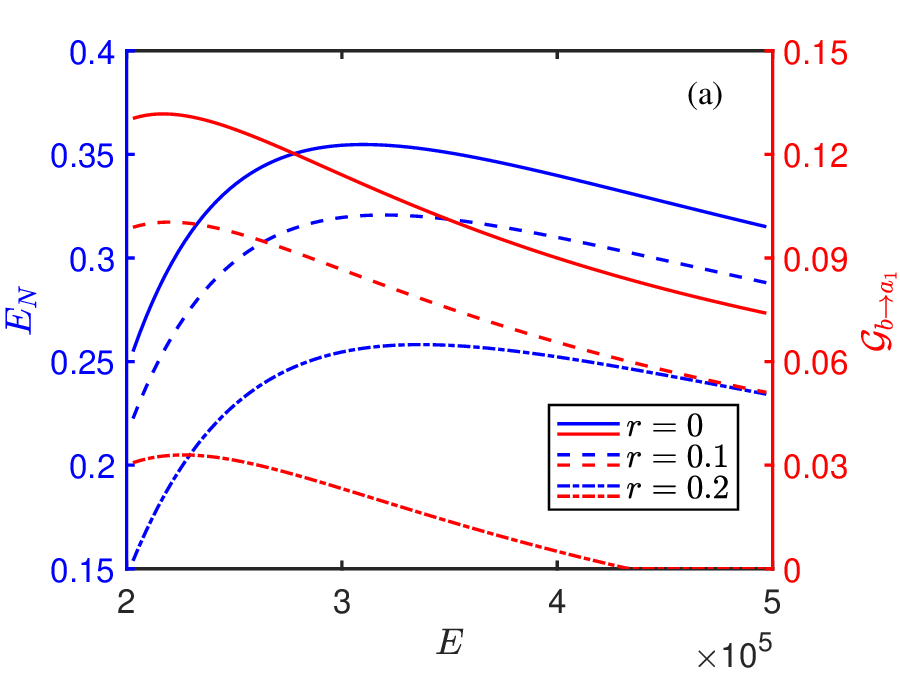}
\includegraphics[width=0.5\columnwidth]{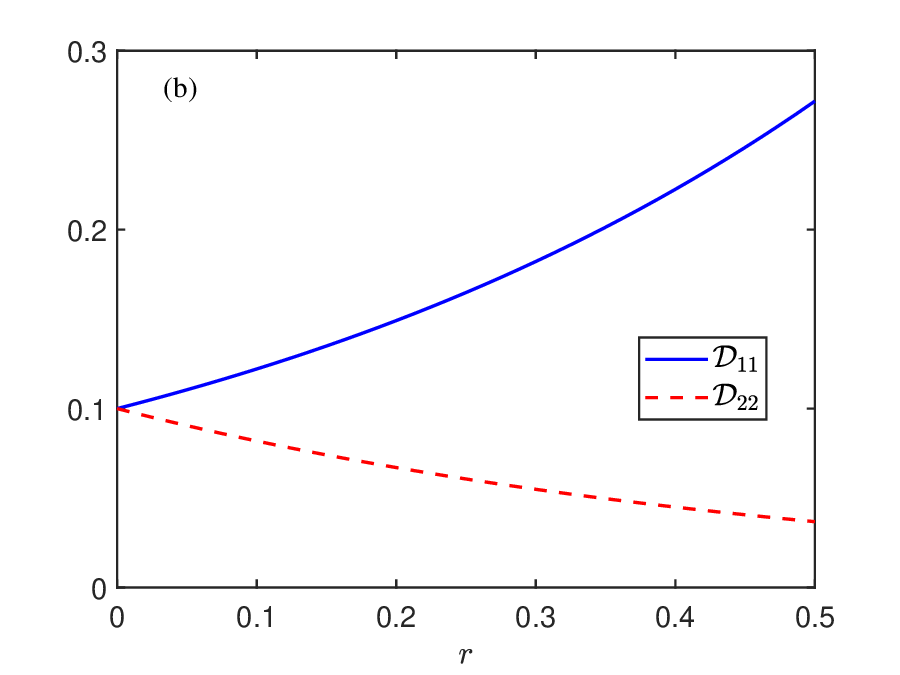}
\caption{(a) The variation in quantum entanglement $E_N$ and EPR steering $\mathcal{G}_{b\rightarrow a_1}$ with pump amplitude $E$ under different squeezing parameters $r$. (b) The behavior  of elements $\mathcal{D}_{11}$ and $\mathcal{D}_{22}$ versus the squeezing parameter $r$. The reference phase is $\theta=0$, and the other parameters are the same as in Fig.\ref{fig2}.}
\label{fig3}
\end{figure}

\section{Effects of the squeezed vacuum field}\label{sec4}
In our model, the squeezed vacuum field's intensity is assumed to be very faint, so it can be dealt as a noise term that only affects the dynamics of the quadrature fluctuation operators. The squeezed vacuum field can be characterized by the squeezing parameter $r$ and reference phase $\theta$. In Fig. \ref{fig3}(a), we display the effects of the squeezing parameter $r$ on the quantum entanglement $E_N$ and EPR steering $\mathcal{G}_{b\rightarrow a_1}$ when the reference phase is $\theta=0$, and the other parameters are chosen the same as given in Fig. \ref{fig2}. The solid lines are $r=0$, the dashed lines correspond to $r=0.1$, and the dash-dotted lines denote $r=0.2$; the quantum entanglement $E_N$ is represented by the blue lines, and the EPR steering $\mathcal{G}_{b\rightarrow a_1}$ is expressed by the red lines. As the pump amplitude $E$ increases, both the quantum entanglement and EPR steering will increase at first and then decrease gradually when the system enters the stable regime, just as the behavior is shown in Fig. \ref{fig2} and explained in Sec. \ref{sec3}. However, both $E_N$ and $\mathcal{G}_{b\rightarrow a_1}$ will weaken with increasing squeezing parameter $r$, which contradicts against the intuition that the squeezed state is beneficial to the quantum process. There are two possible explanations for the physical mechanics behind this.

On the one hand, the squeezed vacuum field input into the WGM mode $a_1$ acts in a similar role as thermal noise according to the noise diffusion matrix $\mathcal{D}$ in Eq. (\ref{eq:zaoshengjuzhen}), and the noise diffusion block matrix $\mathcal{D}_1$ can be simplified as
\begin{align}
\mathcal{D}_{1} & =\frac{\kappa_{1}}{2}\left[2(N+M)+1,0;0,2(N-M)+1\right]
\label{eq:d1huajian}
\end{align}
with the coefficient $M=\sinh r\cosh r$. The elements of $\mathcal{D}_1$, i.e., $\mathcal{D}_{11}=\kappa_{1}(N+M+\frac{1}{2})$ and $\mathcal{D}_{22}=\kappa_{1}(N-M+\frac{1}{2})$, play opposite roles, which are depicted vividly in Fig. \ref{fig3}(b).  The increasing tendency of $\mathcal{D}_{11}$ is more powerful than the decreasing tendency of  $\mathcal{D}_{22}$ under the same $r$. Considering the case that $r\rightarrow\infty$, it is easy to check that the value of element $\mathcal{D}_{11}$ tends to $\infty$; however, the value of element $\mathcal{D}_{22}$ only approaches zero asymptotically. The element $\mathcal{D}_{11}$ heats the environment, and $\mathcal{D}_{22}$ cools the corresponding noise. The competition between the heating and cooling effects determines the dynamical behavior of the covariance matrix $\mathcal{V}$ in Eq. (\ref{eq:xiefangcha}). From the energy point of view, a weak squeezed vacuum field can also be considered another driver of the WGM mode $a_1$, and the comprehensive effects of the pump laser and the squeezed vacuum field can alter the effective pump amplitude of the Hamiltonian (\ref{eq:Hyuanshi}). Therefore, the quantum nonlocality will be influenced simultaneously.

\begin{figure}[t]
\centering
\includegraphics[width=0.5\columnwidth]{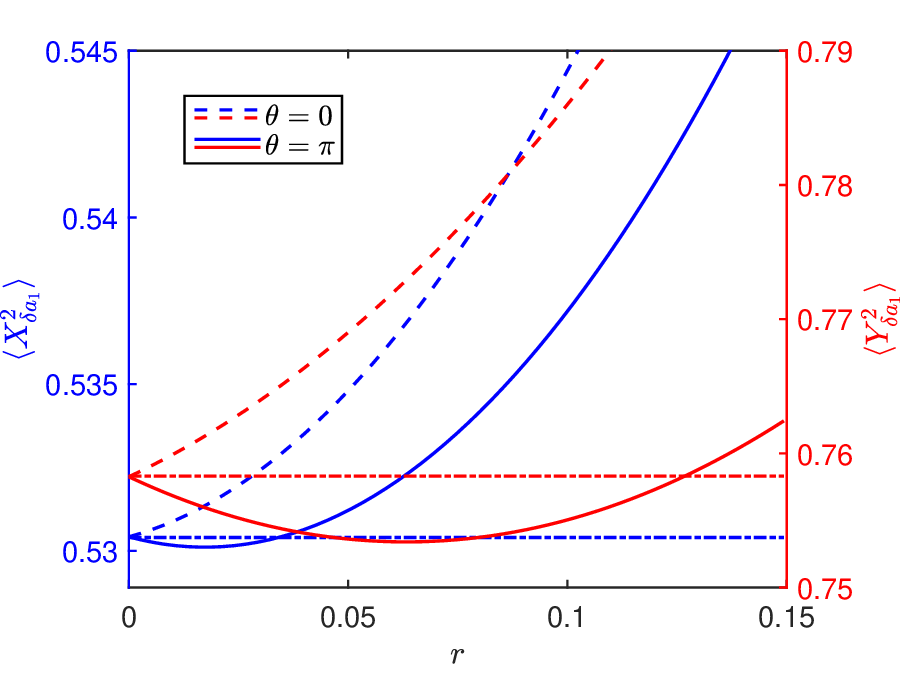}
\caption{The variations in quantum fluctuation operators $\langle X_{\delta a_1}^2\rangle$ and $\langle Y_{\delta a_1}^2\rangle$ versus the squeezing parameter $r$. The system parameters are chosen the same as those in Fig. \ref{fig2}.}
\label{fig4}
\end{figure}

On the other hand, one can inspect the variances of the quadrature fluctuation operators $\langle X^2_{\delta a_1}\rangle$ and  $\langle Y^2_{\delta a_1}\rangle$ of the WGM mode $a_1$ as a function of the squeezing parameter $r$, which are displayed in Fig. \ref{fig4}. The blue lines denote the variance $\langle X^2_{\delta a_1}\rangle$, and the red lines denote the variance $\langle Y^2_{\delta a_1}\rangle$; the dashed lines indicate that the reference phase is $\theta=0$, and the solid lines indicate that the reference phase is $\theta=\pi$. According to Fig. \ref{fig4}, both the variances $\langle X^2_{\delta a_1}\rangle$ and $\langle Y^2_{\delta a_1}\rangle$ will be increased as the squeezing parameter is enhanced because the squeezed state cannot be transferred to the cavity mode effectively when the reference phase is $\theta=0$, which is harmful to the quantum nonlocality. When the reference phase $\theta$ is changed, such as $\theta=\pi$, the variances $\langle X^2_{\delta a_1}\rangle$ and $\langle Y^2_{\delta a_1}\rangle$ become slightly smaller under a certain range of $r$, which are expressed by the value of variances below the corresponding $r=0$. From the perspective of quantum nonlocality, the squeezed state is partially transferred to the cavity mode, so the negative effect induced by the squeezing parameter $r$ is eliminated, and the quantum nonlocality will be enhanced in the same area. Moreover, the noise diffusion subblock matrix $\mathcal{D}_1$ is influenced by the nonzero reference phase $\theta$ and arrives at $\frac{\kappa_{1}}{2}\left[2(N-M)+1,0;0,2(N+M)+1\right]$ with $\theta=\pi$. The dynamics and tendency of its elements $\mathcal{D}_{11}$ and $\mathcal{D}_{22}$ are totally opposite compared with those in Eq. (\ref{eq:d1huajian}). Therefore, one can expect that the reference phase $\theta$ plays a crucial role in enhancing the quantum entanglement and EPR steering when the squeezing parameter $r$ is chosen carefully.

\begin{figure}[t]
\centering
\includegraphics[width=0.5\columnwidth]{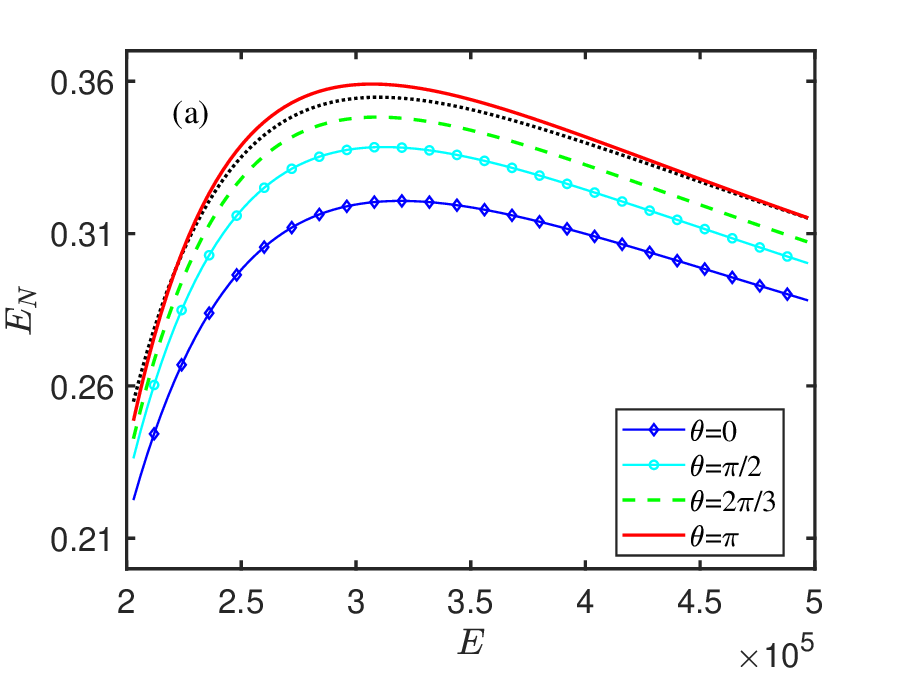}
\includegraphics[width=0.5\columnwidth]{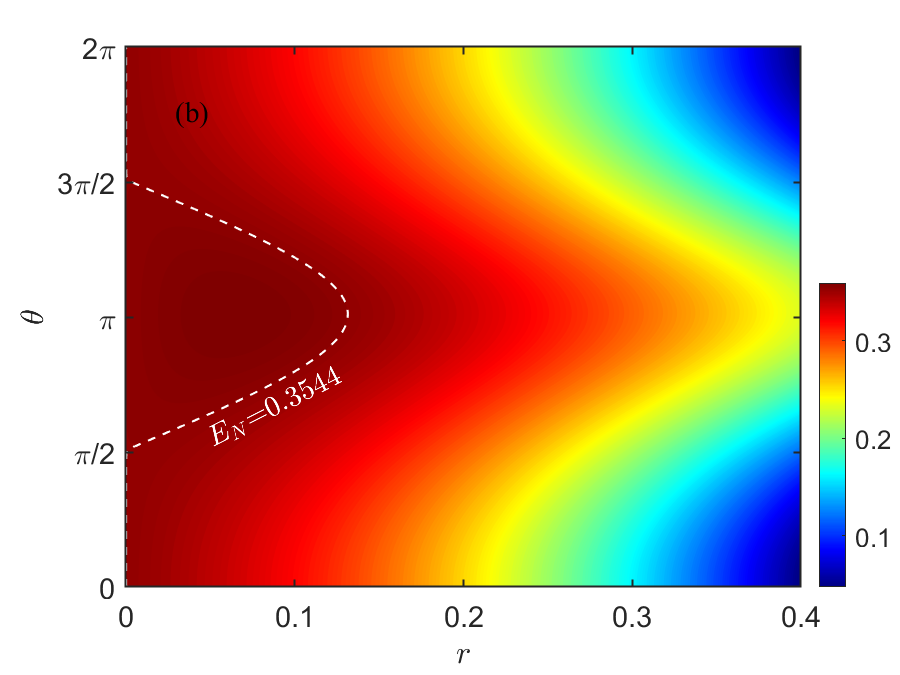}
\includegraphics[width=0.5\columnwidth]{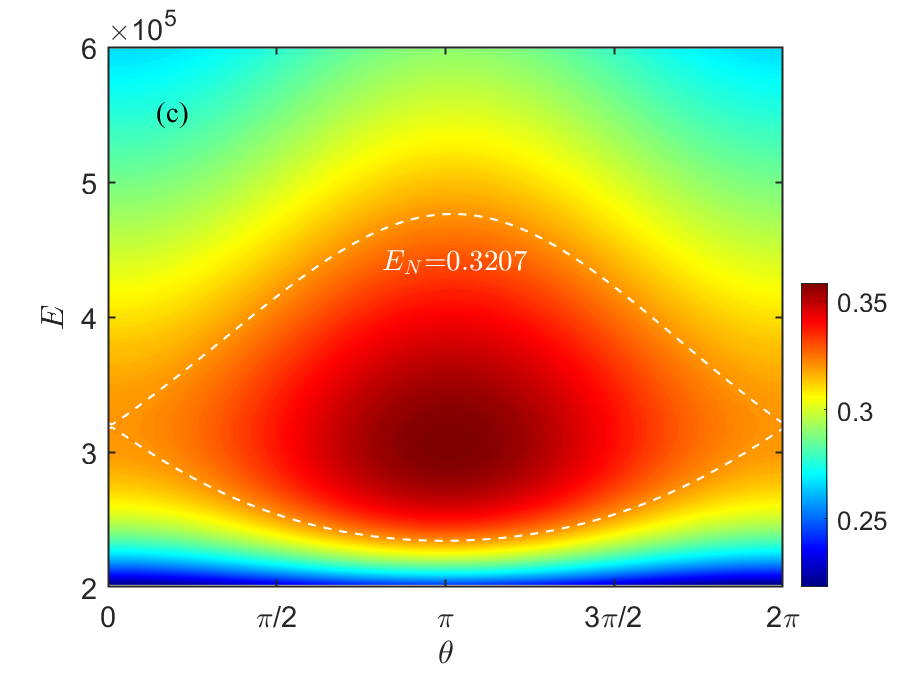}
\caption{(a) Comparison of $E_N$ without/with a squeezed vacuum field. (b) $E_N$ as a function of squeezing parameter $r$ and reference phase $\theta$ under $E=3.0\times10^5$. (c) The variation of $E_N$ with $\theta$ and pump amplitude $E$ under $r=0.1$. The dashed contour lines in (b) and (c) correspond to the quantum entanglement with vacuum noise. The other system parameters are chosen the same as those in Fig. \ref{fig2}.}
\label{fig5}
\end{figure}

Through our numerical analysis, the quantum entanglement $E_N$ and EPR steering $\mathcal{G}_{b\rightarrow a_1}$ display very similar features under the hierarchy of nonlocality. We only exhibit the variations in quantum entanglement $E_N$ under different system parameters in Fig. \ref{fig5}, and the behaviors of EPR steering $\mathcal{G}_{b\rightarrow a_1}$ are not shown here. If not specified, the parameters used here are the same as those in Fig. \ref{fig2}. The effects of the reference phase $\theta$ on the quantum entanglement $E_N$ are shown in panel (a). The dotted black line is the quantum entanglement $E_N$ with vacuum noise, and the other lines are $E_N$ under the squeezed vacuum field ($r=0.1$) with different reference phases $\theta$, where the blue diamond line is $\theta=0$, the cyan circle line is $\theta=\pi/2$, the green dashed line is $\theta=2\pi/3$, and the red line is $\theta=\pi$. The adverse effects induced by the squeezing parameter $r$ are diminished as the reference phase $\theta$ increases from $0$ to $\pi$ and eliminated in the vicinity of $\theta=\pi$, and the quantum entanglement $E_N$ can be enhanced.

\begin{figure}[b]
\centering
\includegraphics[width=0.5\columnwidth]{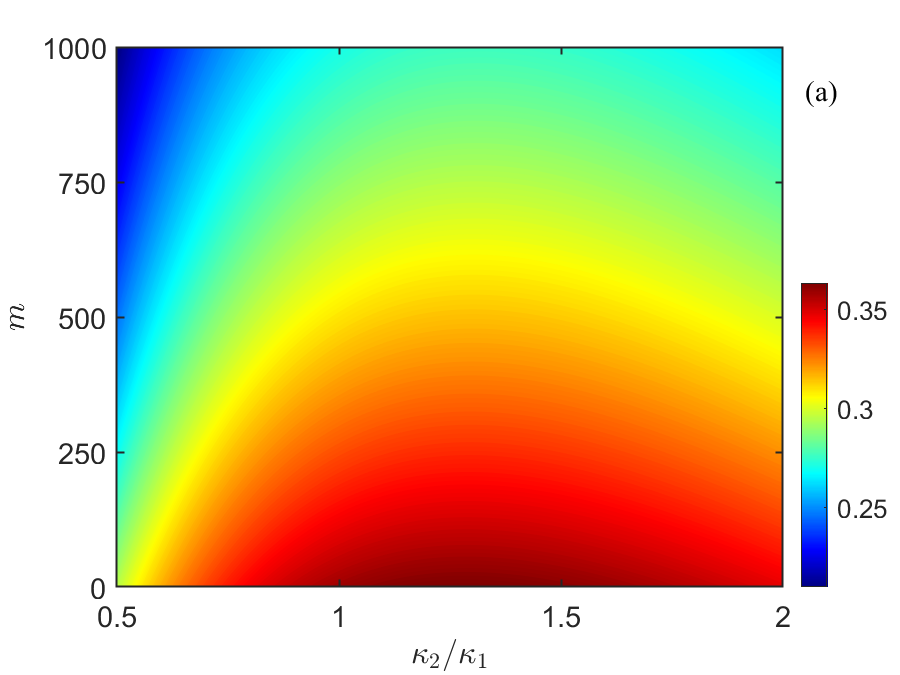}
\includegraphics[width=0.5\columnwidth]{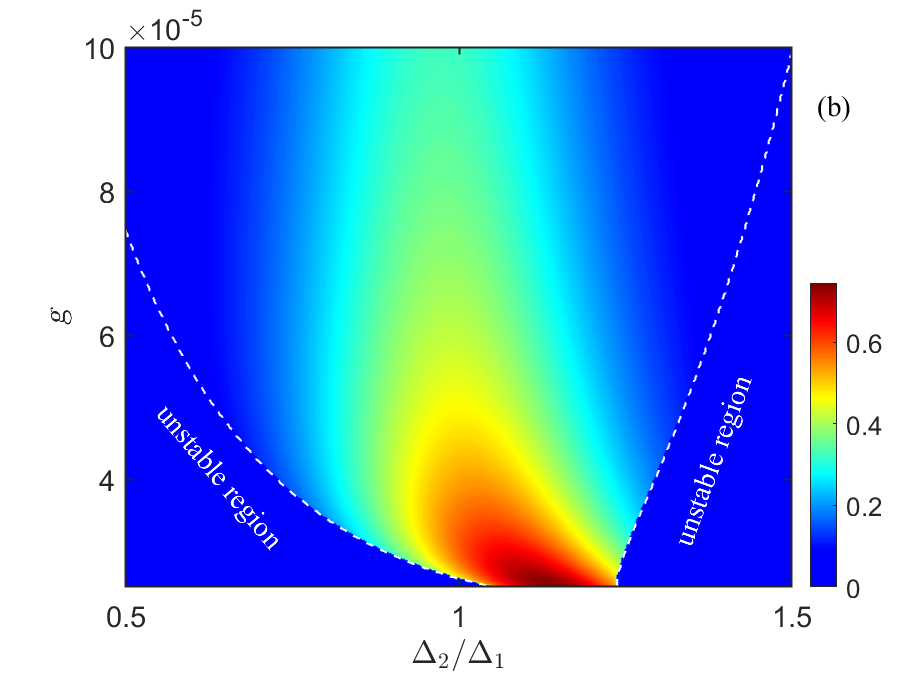}
\caption{(a) $E_N$ versus the ratio of decay rates $\kappa_2/\kappa_1$ and mean thermal phonon number $m$. (b) $E_N$ versus the ratio of detuning $\Delta_2/\Delta_1$ and the single-photon coupling strength $g$ with coupling strength $J=1.0$. The other parameters are the same as given in Figs. \ref{fig2} and \ref{fig5}.}
\label{fig6}
\end{figure}

The variation in $E_N$ as functions of the squeezing parameter $r$ and reference phase $\theta$ is shown in panel (b). The effect of $\theta$ on the quantum entanglement is symmetric about $\theta=\pi$, which is confirmed based on the symmetry analysis of the noise diffusion matrix $\mathcal{D}$ in Eq. (\ref{eq:zaoshengjuzhen}). When the degree of squeezed vacuum is small, the loss of entanglement induced by the squeezing parameter $r$ can be compensated and eliminated by the nonzero reference phase $\theta$, and the quantum entanglement $E_N$ can be enhanced remarkably when $\theta$ is near $\pi$. The quantum entanglement without a squeezed vacuum field is $0.3544$ (the dashed contour line). It can be enhanced where the areas are located inside the contour line, which can be realized by properly controlling the reference phase $\theta$. For a squeezed vacuum field with large $r$, the negative effect on the quantum nonlocality cannot be eliminated by adjusting the reference phase $\theta$. The quantum entanglement will be decreased invariably. In panel (c), we show entanglement enhancement by adjusting the pump amplitude $E$ and reference phase $\theta$ jointly when the squeezing parameter is fixed, such as $r=0.1$. The dashed contour line is $E_N=0.3207$ under the pump amplitude $E\simeq3.2\times10^5$ with the squeezed vacuum field absent. The entanglement is enhanced effectively by adjusting the reference phase and the pump amplitude.

Strong and stable entanglement can be generated and enhanced by chosen system parameters properly assisted by a squeezed vacuum field. In Fig. \ref{fig6}, we show the effects of other system parameters on the entanglement where the squeezing parameters are $r=0.1$ and $\theta=\pi$. The variation in $E_N$ with the ratio of decay rates $\kappa_2/\kappa_1$ and the mean thermal phonon number $m$ is shown in Fig. \ref{fig6}(a). The entanglement $E_N$ consistently decreases with increasing $m$, i.e., the thermal noise of the mechanical mode is monotonically harmful to entanglement. According to the effective Hamiltonian (\ref{eq:linH}) and drift matrix (\ref{eq:shifitmatrix}), the effective detuning $\Delta_1'$ contains the detuning $\Delta_1$ and the displacement part $2g\text{Re}\langle b\rangle$, which will lead to inconsistency between the WGM modes $a_1$ and $a_2$. As displayed in Fig. \ref{fig6}(a), the identical decay rates for WGM modes $a_1$ and $a_2$ are not the optimal choice to generate quantum entanglement. The quantum entanglement $E_N$ as functions of the detuning ratio $\Delta_2/\Delta_1$ and the single-photon optomechanical coupling strength $g$ under the coupling strength $J=1.0$ is displayed in Fig. \ref{fig6}(b). As shown in panel (a), the identical detuning $\Delta_2=\Delta_1$ is not optimal for quantum entanglement. The optimal ratio is associated with $g$, which is determined by the displacement $2g\text{Re}\langle b\rangle$, and the system's stability is significantly influenced by $\Delta_2/\Delta_1$ and $g$, which is similar to that exhibited in Fig. \ref{fig2}.

Moreover, we also investigate the quantum nonlocality between the WGM modes $a_1$ and $a_2$. Because the effects of a squeezed vacuum on the quantum entanglement and EPR steering are similar to those between modes $a_1$ and $b$, their dynamics are not shown here. The single-photon optomechanical coupling strength $g=5.0\times10^{-5}$ is too strong to generate quantum entanglement between WGM modes $a_1$ and $a_2$ \cite{zyl20}. Therefore, the single-photon coupling strength is set as $g=10^{-7}$, and the other system parameters are changed with $g$ simultaneously, for example, $\kappa_1=\kappa_2=0.1$, $J=0.6$, and the stable entanglement can be generated by changing the detuning and pump amplitude. Different from the quantum nonlocality between modes $a_1$ and $b$, EPR steering $\mathcal{G}_{a_1\rightarrow a_2}$ or $\mathcal{G}_{a_2\rightarrow a_1}$ is absent in the above stable regime. Therefore, there is no steerability between these two WGM modes.

\section{Discussion and Conclusions}\label{sec5}
We will discuss the experimental feasibility \cite{pb14,cl14,jh14,sz16,gis10}. The employed pump drive is a 1550-nm laser with frequency $\omega$, the resonance frequencies of WGM modes $a_1$ and $a_2$ are $\omega_{1,2}=2\pi\times193.4\text{THz}$, and their quality factors are chosen to be identical, such as $Q=3\times10^7$, so the decay rates are $\kappa_{1,2}=2\pi\times6.43\text{MHz}$. The coupling strength between modes $a_1$ and $a_2$ can be varied by adjusting the gap between the two resonators and is set as $J=2\pi\times16.1\text{MHz}$. The frequency of the mechanical oscillator is $\omega_m=2\pi\times23.4\text{MHz}$, and the corresponding quality factor can reach $Q_m=10^5$ with the advancement of technology. The pump strength $E=\sqrt{\kappa P_{\text{in}}/\hbar\omega}$ is determined by the power of pump laser and coupling between the pump laser and the WGM resonator, where the power is $P_{\text{in}}\simeq0.2\text{mW}$ for pump strength $E=3.2\times10^5\omega_m$. The squeezed vacuum field can be realized by utilizing a compact monolithic PPKTP cavity \cite{as13}, optical heterodyne sideband modulation locking \cite{gy19}, optical parameter amplification \cite{kt20}, and so on. The squeezing degree is only $0.87\text{dB}$ when the squeezing parameter is $r=0.1$. According to the above analysis, this  project is easily realized under the current experimental conditions.

In conclusion, we have investigated the behavior of the quantum nonlocality (including the quantum entanglement and EPR steering) of coupled optomechanical WGM resonators with a weak squeezed vacuum field when the displacement of detuning induced by the mechanical mode is considered. Strong and stable quantum entanglement and EPR steering between the optical and mechanical modes can be generated in the stable regime by correctly choosing system parameters. When the degree of the squeezed vacuum field is large, it is counterproductive to enhance the quantum nonlocality. However, for a weakly squeezed vacuum field, the reference phase $\theta$ rather than the squeezing parameter $r$ plays an essential role in enhancing the quantum nonlocality compared with the case in which the squeezed vacuum field is absent. Our work can deepen the understanding of the squeezed vacuum field and provide an effective method to enhance the quantum entanglement and EPR steering of optomechanical systems. Furthermore, the effects of the strong intensity of the squeezed vacuum field on the quantum nonlocality of the optomechanical system and the behavior of tripartite entanglement of optomechanical systems under squeezed vacuum field deserve to be investigated.

\begin{backmatter}
\bmsection{Funding}
National Key Research and Development Program of China (2021YFA1402002), National Natural Science Foundation of China (12104277, 12104278, 12175029, 12204440), Fundamental Research Program of Shanxi Province (20210302123063, 202103021223184).

\bmsection{Disclosures}
The authors declare no conflicts of interest.

\bmsection{Data Availability}
The data that support the findings of this study are available upon reasonable request from the authors.
\end{backmatter}

\newcommand{\doi}[2]{\href{https://doi.org/#1}{\color{blue}#2}}


\begin{thebibliography}{99}
\bibitem{ea35} A. Einstein, B. Podolsky, and N. Rosen, ``Can quantum-mechanical description of physical reality be considered complete?'' \doi{10.1103/PhysRev.47.777}{Phys. Rev.} \textbf{47}, 777-780 (1935).
\bibitem{hr09} R. Horodecki, P. Horodecki, M. Horodecki, and K. Horodecki, ``Quantum entanglement,''  \doi{10.1103/RevModPhys.81.865}{Rev. Mod. Phys.} \textbf{81}, 865-942 (2009).
\bibitem{okcf18}  C. F. Ockeloen-Korppi, E. Damsk\"{a}gg, J. M. Pirkkalainen, M. Asjad, A. A. Clerk, F. Massel, M. J. Woolley, and M. A. Sillanp\"{a}\"{a}, ``Stabilized entanglement of massive mechanical oscillators,''  \doi{10.1038/s41586-018-0038-x}{Nature} \textbf{556}, 478-482 (2018).
\bibitem{se35} E. Schr\"{o}dinger, ``Discussion of probability relations between separated systems,''  \doi{10.1017/S0305004100013554}{Math. Proc. Cambridge Philos. Soc.} \textbf{31}, 555-563 (1935).
\bibitem{rmd09} M. D. Reid, P. D. Drummond, W. P. Bowen, E. G. Cavalcanti, P. K. Lam, H. A. Bachor, U. L. Andersen, and G. Leuchs, ``Colloquium: The Einstein-Podolsky-Rosen paradox: from concepts to applications,'' \doi{10.1103/RevModPhys.81.1727}{Rev. Mod. Phys.} \textbf{81}, 1727-1751 (2009).
\bibitem{rmd89} M. D. Reid,  ``Demonstration of the Einstein-Podolsky-Rosen paradox using nondegenerate parametric amplification,'' \doi{/10.1103/PhysRevA.40.913}{Phys. Rev. A} \textbf{40}, 913-923 (1989).
\bibitem{wsp11} S. P. Walborn, A. Salles, R. M. Gomes, F. Toscano, and P. H. Souto Ribeiro, ``Revealing hidden Einstein-Podolsky-Rosen nonlocality,'' \doi{10.1103/PhysRevLett.106.130402}{Phys. Rev. Lett.} \textbf{106}, 130402 (2011).
\bibitem{am14} M. Aspelmeyer, T. J. Kippenberg, and F. Marquardt, ``Cavity optomechanics,''  \doi{10.1103/RevModPhys.86.1391}{Rev. Mod. Phys.} \textbf{86}, 1391-1452 (2014).
\bibitem{mi18} I. Marinkovi\'{c}, A. Wallucks, R. Riedinger, S. Hong, M. Aspelmeyer, and S. Gr\"{o}blacher, ``Optomechanical Bell test,''  \doi{10.1103/PhysRevLett.121.220404}{Phys. Rev. Lett.} \textbf{121}, 220404 (2018).
\bibitem{vd07} D. Vitali, S. Gigan, A. Ferreira, H. R. B\"{o}hm, P. Tombesi, A. Guerreiro, V. Vedral, A. Zeilinger, and M. Aspelmeyer, ``Optomechanical entanglement between a movable mirror and a cavity field,'' \doi{10.1103/PhysRevLett.98.030405}{Phys. Rev. Lett.} \textbf{98}, 030405 (2007).
\bibitem{gc08} C. Genes, D. Vitali, and P. Tombesi, ``Emergence of atom-light-mirror entanglement inside an optical cavity,'' \doi{10.1103/PhysRevA.77.050307}{Phys. Rev. A} \textbf{77}, 050307(R) (2008).
\bibitem{ma09} A. Mari, and J. Eisert, ``Gently modulating optomechanical systems,'' \doi{10.1103/PhysRevLett.103.213603}{Phys. Rev. Lett.} \textbf{103}, 213603 (2009).
\bibitem{wyd13} Y. D. Wang, and A. A. Clerk, ``Reservoir-engineered entanglement in optomechanical systems,'' \doi{10.1103/PhysRevLett.110.253601}{Phys. Rev. Lett.} \textbf{110}, 253601 (2013).
\bibitem{hx15} X. Hu, ``Entanglement generation by dissipation in or beyond dark resonances,'' \doi{10.1103/PhysRevA.92.022329}{Phys. Rev. A} \textbf{92}, 022329 (2015).
\bibitem{cs17} S. Chakraborty, and A. K. Sarma, ``Enhancing quantum correlations in an optomechanical system via cross-Kerr nonlinearity,''  \doi{10.1364/JOSAB.34.001503}{J. Opt. Soc. Am. B} \textbf{34}, 1503-1510 (2017).
\bibitem{hcs20} C. S. Hu, Z. Q. Liu, Y. Liu, L. T. Shen, H. Wu, and S. B. Zheng, ``Entanglement beating in a cavity optomechanical system under two-field driving,'' \doi{10.1103/PhysRevA.101.033810}{Phys. Rev. A} \textbf{101}, 033810 (2020).
\bibitem{ldg21} D. G. Lai, W. Qin, B. P. Hou, A. Miranowicz and F. Nori, ``Significant enhancement in refrigeration and entanglement in auxiliary-cavity-assisted optomechanical systems,'' \doi{10.1103/PhysRevA.104.043521}{Phys. Rev. A} \textbf{104}, 043521 (2021).
\bibitem{hxz21} X. Z. Hao, X. Y. Zhang, Y. H. Zhou, W. Li, S. C. Hou, and X. X. Yi, ``Dynamical bipartite and tripartite entanglement of mechanical oscillators in an optomechanical array,'' \doi{10.1103/PhysRevA.104.053515}{Phys. Rev. A} \textbf{104}, 053515 (2021).
\bibitem{zwj21} W. J. Zhang, Y. Zhang, Q. Guo, A. P. Liu, G. Li, and T. Zhang, ``Strong mechanical squeezing and optomechanical entanglement in a dissipative double-cavity system via pump modulation,'' \doi{10.1103/PhysRevA.104.053506}{Phys. Rev. A} \textbf{104}, 053506 (2021).
\bibitem{bch21} C. H. Bai, D. Y. Wang, S. Zhang, S. Liu, and H. F. Wang, ``Generation of strong mechanical entanglement by pump modulation,''  \doi{10.1002/qute.202000149}{Adv. Quantum Technol.} \textbf{4}, 2000149 (2021).
\bibitem{wf22} F. Wang, C. Gou, J. Xu, and C. Gong, ``Hybrid magnon-atom entanglement and magnon blockade via quantum interference,'' \doi{10.1103/PhysRevA.106.013705}{Phys. Rev. A} \textbf{106}, 013705 (2022).
\bibitem{ljq11} J. Q. Liao, and C. K. Law, ``Parametric generation of quadrature squeezing of mirrors in cavity optomechanics,''  \doi{10.1103/PhysRevA.83.033820}{Phys. Rev. A} \textbf{83}, 033820 (2011).
\bibitem{lyc13} Y. C. Liu, Y. F. Xiao, X. Luan, and C. W. Wong, ``Dynamic dissipative cooling of a mechanical resonator in strong coupling optomechanics,'' \doi{10.1103/PhysRevLett.110.153606}{Phys. Rev. Lett.} \textbf{110}, 153606 (2013).
\bibitem{gy14} Y. Guo, K. Li, W. Nie, and Y. Li, ``Electromagnetically-induced-transparency-like ground-state cooling in a double-cavity optomechanical system,'' \doi{10.1103/PhysRevA.90.053841}{Phys. Rev. A} \textbf{90}, 053841 (2014).
\bibitem{lxy15pra} X. Y. L\"{u}, J. Q. Liao, L. Tian, and F. Nori, ``Steady-state mechanical squeezing in an optomechanical system via Duffing nonlinearity,'' \doi{10.1103/PhysRevA.91.013834}{Phys. Rev. A} \textbf{91}, 013834 (2015).
\bibitem{hx19} X. Han, D. Y. Wang, C. H. Bai, W. X. Cui, S. Zhang, and H. F. Wang, ``Mechanical squeezing beyond resolved sideband and weak-coupling limits with frequency modulation,'' \doi{10.1103/PhysRevA.100.033812}{Phys. Rev. A} \textbf{100}, 033812 (2019).
\bibitem{xb20} B. Xiong, X. Li, S. L. Chao, Z. Yang, W. Z. Zhang, W. Zhang, and L. Zhou, ``Strong mechanical squeezing in an optomechanical system based on Lyapunov control,'' \doi{10.1364/PRJ.8.000151}{Photon. Res.} \textbf{8}, 151-159 (2020).
\bibitem{bch20} C. H. Bai, D. Y. Wang, S. Zhang, S. Liu, and H. F. Wang, ``Strong mechanical squeezing in a standard optomechanical system by pump modulation,'' \doi{10.1103/PhysRevA.101.053836}{Phys. Rev. A} \textbf{101}, 053836 (2020).
\bibitem{hq15} Q. He, L. Rosales-Z\'{a}rate, G. Adesso, and M. D. Reid, ``Secure continuous variable teleportation and Einstein-Podolsky-Rosen steering,'' \doi{10.1103/PhysRevLett.115.180502}{Phys. Rev. Lett.} \textbf{115}, 180502 (2015).
\bibitem{sfx17} F. X. Sun, D. Mao, Y. T. Dai, Z. Ficek, Q. Y. He, and Q. H. Gong, ``Phase control of entanglement and quantum steering in a three-mode optomechanical system,'' \doi{10.1088/1367-2630/aa9c9a}{New J. Phys.} \textbf{19}, 123039 (2017).
\bibitem{zj15} J. Zhang, T. Zhang, A. Xuereb, D. Vitali, and J. Li, ``More nonlocality with less entanglement in a tripartite atom-optomechanical system,'' \doi{10.1002/andp.201400107}{Ann. Phys. (Berlin)} \textbf{527}, 147-155 (2015).
\bibitem{th15} H. Tan, X. Zhang, and G. Li, ``Steady-state one-way Einstein-Podolsky-Rosen steering in optomechanical interfaces,''  \doi{10.1103/PhysRevA.91.032121}{Phys. Rev. A} \textbf{91}, 032121 (2015).
\bibitem{ps07} S. Pielawa, G. Morigi, D. Vitali, and L. Davidovich,  ``Generation of Einstein-Podolsky-Rosen-Entangled radiation through an atomic reservoir,'' \doi{10.1103/PhysRevLett.98.240401}{Phys. Rev. Lett.} \textbf{98}, 240401 (2007).
\bibitem{zw17} W. Zhong, G. Cheng, and X. Hu, ``One-way Einstein-Podolsky-Rosen steering via atomic coherence,'' \doi{10.1364/OE.25.011584}{Opt. Express} \textbf{25}, 11584-11597 (2017).
\bibitem{gq23} Q. Guo, M. R. Wei, C. H. Bai, Y. Zhang, G. Li, and T. Zhang, ``Manipulation and enhancement of Einstein-Podolsky-Rosen steering between two mechanical modes generated by two Bogoliubov dissipation pathways,'' \doi{10.1103/PhysRevResearch.5.013073}{Phys. Rev. Res.} \textbf{5}, 013073 (2023).
\bibitem{lcg20} C. G. Liao, H. Xie, R. X. Chen, M. Y. Ye, and X. M. Lin, ``Controlling one-way quantum steering in a modulated optomechanical system,'' \doi{10.1103/PhysRevA.101.032120}{Phys. Rev. A} \textbf{101}, 032120 (2020).
\bibitem{zs19} S. Zheng, F. Sun, Y. Lai, Q. Gong, and Q. He, ``Manipulation and enhancement of asymmetric steering via interference effects induced by closed-loop coupling,'' \doi{10.1103/PhysRevA.99.022335}{Phys. Rev. A} \textbf{99}, 022335 (2019).
\bibitem{lj17} J. Li, G. Li, S. Zippilli, D. Vitali, and T. Zhang, ``Enhanced entanglement of two different mechanical resonators via coherent feedback,'' \doi{10.1103/PhysRevA.95.043819}{Phys. Rev. A} \textbf{95}, 043819 (2017).
\bibitem{zw22} W. Zhang, T. Wang, X. Han, S. Zhang, and H. F. Wang, ``Quantum entanglement and one-way steering in a cavity magnomechanical system via a squeezed vacuum field,'' \doi{10.1364/OE.453787}{Opt. Express} \textbf{30}, 10969-10980 (2022).
\bibitem{yzb23} Z. B. Yang, Y. Ming, R. C. Yang, And H. Y. Liu, ``Asymmetric transmission and entanglement in a double-cavity magnomechanical system,'' \doi{10.1364/JOSAB.481012}{J. Opt. Soc. Am. B}, \textbf{40}, 822-829 (2023).
\bibitem{lxy15} X. Y. L\"{u}, Y. Wu, J. R. Johansson, H. Jing, J. Zhang, and F. Nori, ``Squeezed optomechanics with phase-matched amplification and dissipation,'' \doi{10.1103/PhysRevLett.114.093602}{Phys. Rev. Lett.} \textbf{114}, 093602 (2015).
\bibitem{tl22} L. Tang, J. Tang, M. Chen, F. Nori, M. Xiao, and K. Xia, ``Quantum squeezing induced optical nonreciprocity,'' \doi{10.1103/PhysRevLett.128.083604}{Phys. Rev. Lett.} \textbf{128}, 083604 (2022).
\bibitem{zcj20} C. J. Zhu, L. L. Ping, Y. P. Yang, and G. S. Agarwal, ``Squeezed light induced symmetry breaking superradiant phase transition,'' \doi{10.1103/PhysRevLett.124.073602}{Phys. Rev. Lett.} \textbf{124}, 073602 (2020).
\bibitem{zw20} W. Zhao, S. D. Zhang, A. Miranowicz, and H. Jing, ``Weak-force sensing with squeezed optomechanics,'' \doi{10.1007/s11433-019-9451-3}{Sci. China-Phys. Mech. Astron.} \textbf{63}, 224211 (2020).
\bibitem{gcw00} C. W. Gardiner, and P. Zoller, Quantum noise, 2nd ed. (Berlin, Springer, 2000).
\bibitem{smo97} M. O. Scully, and M. S. Zubairy, Quantum optics. (Cambridge, Cambridge University Press, 1997).
\bibitem{vg02} G. Vidal, and R. F. Werner, ``Computable measure of entanglement,'' \doi{10.1103/PhysRevA.65.032314}{Phys. Rev. A} \textbf{65}, 032314 (2002).
\bibitem{ag04} G. Adesso, A. Serafini, and F. Illuminati, ``Extremal entanglement and mixedness in continuous variable systems,'' \doi{10.1103/PhysRevA.70.022318}{Phys. Rev. A} \textbf{70}, 022318 (2004).
\bibitem{pmb05} M. B. Plenio, ``Logarithmic Negativity: a full entanglement monotone that is not convex,'' \doi{10.1103/PhysRevLett.95.090503}{Phys. Rev. Lett.} \textbf{95}, 090503 (2005).
\bibitem{ki15} I. Kogias, A. R. Lee, S. Ragy, and G. Adesso, ``Quantification of Gaussian quantum steering,'' \doi{10.1103/PhysRevLett.114.060403}{Phys. Rev. Lett.} \textbf{114}, 060403 (2015).
\bibitem{whm07} H. M. Wiseman, S. J. Jones, and A. C. Doherty, ``Steering, entanglement, nonlocality, and the Einstein-Podolsky-Rosen paradox,'' \doi{10.1103/PhysRevLett.98.140402}{Phys. Rev. Lett.} \textbf{98}, 140402 (2007).
\bibitem{dex87} E. X. DeJesus, and C. Kaufman, ``Routh-Hurwitz criterion in the examination of eigenvalues of a system of nonlinear ordinary differential equations,'' \doi{10.1103/PhysRevA.35.5288}{Phys. Rev. A} \textbf{35}, 5288-5290 (1987).
\bibitem{zyl20} Y. L. Zhang, C. S. Yang, Z. Shen, C. H. Dong, G. C. Guo, C. L. Zou, and X. B. Zou, ``Enhanced optomechanical entanglement and cooling via dissipation engineering,'' \doi{10.1103/PhysRevA.101.063836}{Phys. Rev. A} \textbf{101}, 063836 (2020).
\bibitem{pb14} B. Peng, S. K. \"{O}zdemir, F. Lei, F. Monifi, M. Gianfreda, G. L. Long, S. Fan, F. Nori, C. M. Bender, and L. Yang, ``Parity-time-symmetric whispering-gallery microcavities,'' \doi{10.1038/NPHYS2927}{Nature Phys.} \textbf{10}, 394-398 (2014).
\bibitem{cl14} L. Chang, X. Jiang, S. Hua, C. Yang, J. Wen, L. Jiang, G. Li, G. Wang, and M. Xiao, ``Parity-time symmetry and variable optical isolation in active-passive-coupled microresonators,'' \doi{10.1038/NPHOTON.2014.133}{Nature Photon.} \textbf{8}, 524-529 (2014).
\bibitem{sz16} Z. Shen, Y. L. Zhang, Y. Chen, C. L. Zou, Y. F. Xiao, X. B. Zou, F. W. Sun, G. C. Guo, and C. H. Dong, ``Experimental realization of optomechanically induced non-reciprocity,'' \doi{10.1038/NPHOTON.2016.161}{Nature Photon.} \textbf{10}, 657-661 (2016).
\bibitem{gis10} I. S. Grudinin, H. Lee, O. Painter, and K. J. Vahala, ``Phonon laser Aaction in a tunable two-level system,'' \doi{10.1103/PhysRevLett.104.083901}{Phys. Rev. Lett.} \textbf{104}, 083901 (2010).
\bibitem{jh14} H. Jing, S. K. \"{O}zdemir, X. Y. L\"{u}, J. Zhang, L. Yang, and F. Nori, ``$\mathcal{PT}$-symmetric phonon laser,'' \doi{10.1103/PhysRevLett.113.053604}{Phys. Rev. Lett.} \textbf{113}, 053604 (2014).
\bibitem{as13} S. Ast, M. Mehmet, and R. Schnabel, ``High-bandwidth squeezed light at 1550 nm from a compact monolithic PPKTP cavity,'' \doi{10.1364/OE.21.013572}{Opt. Express} \textbf{21}, 13572-13579 (2013).
\bibitem{gy19} Y. Gao, J. Feng, Y. Li, and K. Zhang, ``Generation of stable, squeezed vacuum states at audio frequency using optical serrodyne sideband modulation locking method,'' \doi{10.1088/1612-202X/ab0a60}{Laser Phys. Lett.} \textbf{16}, 055202 (2019).
\bibitem{kt20} T. Kashiwazaki, N. Takanashi, T. Yamashima, T. Kazama, K. Enbutsu, R. Kasahara, T. Umeki, and A. Furusawa, ``Continuous-wave 6-dB-squeezed light with 2.5-THz-bandwidth from single-mode PPLN waveguide,'' \doi{10.1063/1.5142437}{APL Photon.} \textbf{5}, 036104 (2020).
\end{thebibliography}
\end{document}